\documentclass[fleqn,twoside]{article}
\usepackage{espcrc2}

\usepackage{epsfig}

\newcommand{\AmS}{{\protect\the\textfont2
  A\kern-.1667em\lower.5ex\hbox{M}\kern-.125emS}}

\title{The Radiative Return: a review of experimental results}

\author{Achim Denig\address[IEKP]{Universit\"at Karlsruhe, Institut f\"ur Experimentelle Kernphysik\\
                                  Postfach 3640, D-76021 Karlsruhe, Germany}
}
       
\begin{document}

\begin{abstract}
The radiative return is a new method for hadronic cross section measurements 
at electron-positron colliders, which are operated at a fixed center-of-mass energy
(so-called particle factories). In order to lower the effective hadronic mass 
${\rm M}_{\rm hadr}$ only such events are taken, in which one of the electrons or positrons
has emitted an initial state radiation photon. We present precision measurements of
the pion formfactor from the Frascati $\phi$-factory DA$\Phi$NE with the KLOE experiment
and measurements of higher particle multiplicities as well as a measurement of the
timelike proton-antiproton formfactor from the BaBar experiment at the B-factory PEP-II.
These raditative return measurements are compared to results, which are 
obtained by means of an energy scan, i.e. by means of a systematic variation of the
beam energy of the collider. We also report on the impact of
these measurements on the hadronic contribution of the 
anomalous magnetic moment of the muon, which is obtained via a dispersion
integral using hadronic cross section data as input. 
\vspace{1pc}
\end{abstract}

\maketitle

\section{THE RADIATIVE RETURN AND ITS CONNECTION TO THE MUON ANOMALY}
\subsection{The radiative return method}
Modern particle factories, such as the Frascati $\phi$-factory 
DA$\Phi$NE or the B-factories PEP-II and KEK-B are designed for
a fixed center-of-mass energy $\sqrt{s}$. An energy scan for the measurement
of hadronic cross sections is therefore not feasible. A new and
complementary ansatz is given by the 'radiative return' method,
which utilizes events, in which one of the electrons (positrons)
has emitted a photon. The invariant mass $M_{\rm hadr}$ 
of the hadronic system is reduced in this way and the hadronic cross
section in the energy range $M_{\rm hadr}<\sqrt{s}$ becomes
accessible. In order to deduce from the measured {\it radiative} cross section
$\sigma(e^+e^-\to {\rm hadrons}+\gamma)$ the {\it non-radiative} 
cross section $\sigma(e^+e^-\to {\rm hadrons})$, a theoretical 
radiator function is used via the relation:
\begin{eqnarray*}
M^2_{\rm hadr} \cdot \frac{d\sigma_{{\rm hadrons}+\gamma}}{dM^2_{\rm hadr}}=
\sigma_{\rm hadr} \cdot H(M^2_{\rm hadr}) 
\end{eqnarray*}
The radiator function, which describes the initial state radiation (ISR) process,
has been computed up to NLO by the Monte-Carlo generator PHOKHARA~\cite{phok} with
a precision of $0.5\%$. If the measured radiative cross section is 
divided by the radiative muon cross section $e^+e^-\to \mu^+\mu^-\gamma$, the ratio 
$R=\sigma(e^+e^-\to {\rm hadrons})/\sigma(e^+e^-\to \mu^+\mu^-)$ can
be deduced directly. In this case one does not need to normalize the data
to the integrated luminosity, nor is the radiator function needed. However,
an excellent separation of hadrons from muons is required.
\\
In the following we present results from the KLOE 
experiment at the Frascati collider DA$\Phi$NE, which
is operated on the $\phi(1020)$-resonance and allows to measure the 
pion formfactor below 1 GeV. As we will see in the next subchapter, this measurement
is of crucial importance for the hadronic contribution to the anomalous magnetic
moment of the muon and a precision $\simeq 1\%$ or better is needed. 
After that we discuss a series of results from the BaBar-experiment 
for the timelike proton-antiproton formfactor and for several exclusive final states with higher
multiplicities in the mass range from threshold up to $4.5$ GeV. 
The combination of KLOE and BaBar data allows to cover the hadronic cross
section in the entire mass range below $\sim 4.5$ GeV. This is the relevant
energy region needed for a significantly improved determination of the anomalous magnetic
moment of the muon.\\
\subsection{The anomalous magnetic moment of the muon}
\begin{figure}[t]
\epsfig{file=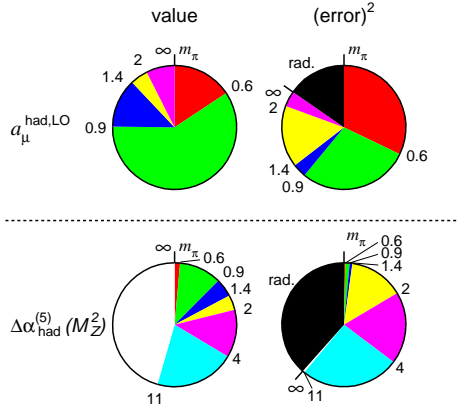, width=7.5cm}
\vspace{-2.5cm}
\caption{\small{The contribution of different energy ranges of hadronic cross
section data to the absolute value and the squared error for 
$a_\mu^{\rm hadr}$ and $\Delta\alpha_{\rm hadr}^{(5)}(M_Z^2)$.}}
\label{pie}
\end{figure}
The recent measurement of the muon anomaly $a_\mu$ 
at the Brookhaven National Laboratory with a precision of $0.5$ ppm~\cite{brookh}
has led to renewed interest in accurate measurements of the cross
section for $e^+e^-$ annihilation into hadrons. 
Hadronic contributions 
to the photon spectral functions due to quark loops 
are not calculable in the framework of perturbative QCD. 
It is well known, however, that the 
hadronic piece of the spectral function 
is connected by unitarity to the cross section for $e^+e^-\to$ hadrons. 
A dispersion relation can thus be derived,
giving the contribution to $a_\mu$ as an integral over the hadronic 
cross section.\\
Also the hadronic contribution to the running of the electromagnetic 
fine structure constant at the $Z$-pole $\Delta\alpha_{\rm hadr}^{(5)}(s)$ 
is not calculable within perturbative QCD (pQCD) at low energies and 
is obtained in a similar way 
by means of a dispersion relation. Current 
electroweak precision tests at the scale $s=M_Z^2$ are 
unfortunately limited by the precision of the order of $1\%$
by which the hadronic piece to the running of $\alpha$ is known. \\ 
Fig.~\ref{pie} (from ref.~\cite{Hagiwara:2003da}) shows the
contributions of different energy intervals of hadronic cross section data 
to the absolute value and to the squared error for $a_\mu^{\rm hadr}$
and $\Delta\alpha_{\rm hadr}^{(5)}(M_Z^2)$. We see that in the
case of the muon anomaly, the radiative return program described above
covers the entire energy range of interest. 
The two-pion cross section (pion formfactor) is 
the by far dominating channel, since below $1$ GeV it 
contributes to $\sim 60\%$ to  $a_\mu^{\rm hadr}$. 
In the case of the fine structure constant $\sim 30\%$ of the absolute 
value of $\Delta\alpha_{\rm hadr}^{(5)}(M_Z^2)$ 
are covered by the mass range $< 4.5$ GeV. 
If pQCD is applicable down to lower energies or in case
novel theoretical techniques can be used in the evaluation of the dispersion
integral, the impact of
the hadronic cross section data below $4.5$ GeV is very much enhanced.
For further details concerning the evaluation of
the dispersion integrals we refer to ref.~\cite{ej}~\cite{dh}~\cite{Hagiwara:2003da}.

\section{THE KLOE MEASUREMENT OF THE PION FORMFACTOR}
KLOE has published a radiative-return analysis of the pion formfactor, 
in which the ISR-photon is required to be emitted at small (large) polar angles 
$\theta_\gamma<15^{\rm o}$ ($\theta_\gamma>165^{\rm o}$) \cite{kloe}. No photon tagging
is possible in such an approach. The high momentum resolution of the KLOE
drift chamber allows a separation of signal from background with very
high precision even without the kinematic closure of the event.
A new and complementary analysis
at large photon angles ($50^{\rm o}<\theta_\gamma<130^{\rm o}$) and
with tagged photons
is under study now and will allow to cover the
threshold region, which was kinematically forbidden before. Moreover,
the beforementioned normalization to radiative muon pairs will allow
an important cross-check of the radiator function approach.
\\
\subsection{Small angle analysis}
The small angle analysis provides high statistics for ISR-photons and 
suppresses the relative amount of events, in which the photon is emitted
from the final state pions (final state radiation, FSR). FSR is an irreducible 
background to the radiative return analysis. The relative amount of 
FSR-events is well below $1\%$ for the chosen selection cuts and has
been estimated by means of the Monte-Carlo-generator PHOKHARA, which uses
the model of scalar QED (pointlike pions) as a model for FSR. Special
attention has been given to events, in which simultaneously an ISR- and
FSR-photon are emitted (NLO-FSR) since those events must not be considered as a background
and are also not suppressed by the acceptance cuts.
The relative contribution of NLO-FSR events is known with a precision of $0.3\%$. 
\\
A total experimental error of $0.9\%$ has been achieved
for the $e^+e^- \to \pi^+\pi^-\gamma$ cross section measurement.
The error consists of the individual
contributions of the selection efficiencies and of the precision,
with which the residual background after all selection cuts is known;
further details can be found in ref.~\cite{kloe}.
The pion formfactor, which can be obtained from the non-radiative
cross section $e^+e^- \to \pi^+\pi^-$, is extracted with a total error of
$1.3\%$. This includes also the
theory uncertainties associated with the radiator function and with the 
large-angle Bhabha cross section.
The Bhabha cross section is needed for the luminosity measurement
and will diminish with the 
new version of the BABAYAGA event generator, see ref.~\cite{babayaga}.
\\
\\
The KLOE measurement of the pion formfactor has been used to compute
the contribution of the two-pion cross section to $a_\mu^{\rm hadr}$ 
in the energy interval $0.35<M_{\pi\pi}^2<0.95$ GeV$^2$. The value
$a_\mu^{\pi\pi}=(388.7 \pm 0.8_{\rm stat} \pm 3.5_{\rm syst} \pm 3.5_{\rm theo}){\rm x}10^{-10}$
covers about $60\%$ of the total contribution. The KLOE result
agrees within $0.5$ standard deviations~\footnote{in the somewhat smaller mass
mass region $0.37<M_{\pi\pi}^2<0.93$ GeV$^2$} 
with values for $a_\mu^{\pi\pi}$ computed from the data sets of the experiments
CMD-2~\cite{cmd2} and SND~\cite{snd2}, which were operated in the last years at the VEPP-2M-collider in Novosibirsk.
The relative difference of the mass spectra is shown in fig.~\ref{diff}. For this comparison the
KLOE data points have been interpolated and the measured data points of CMD-2 and
SND are used in the plot. We observe relatively large deviations of up to some 
percent between KLOE and SND at high and low masses, while the overall agreement
is better with CMD-2. Please notice that SND very recently 
has updated~\cite{snd1} its previous measurement ref.~\cite{snd2},
taking into account a new treatment of radiative corrections. The disagreement with KLOE
has diminished with new data, but a clear trend of difference is still visible. 
The good agreement in the dispersion integral is partly due to a compensation effect
at lower and higher energies.
\\
\begin{figure}[t]
\epsfig{file=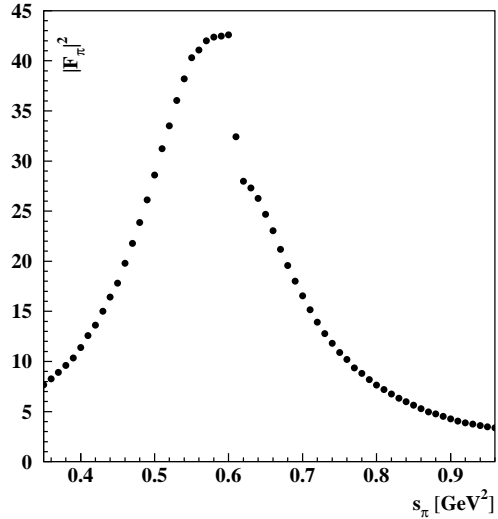, width=7.5cm}
\vspace{-0.75cm}
\caption{\small{KLOE measurement of the pion formfactor using radiative-return events with
photons emitted at small polar angles.}}
\label{fpi}
\end{figure}
\\
The KLOE measurement presented above refers to data taken in 2001 with a total 
integrated luminosity of $\sim 140$ pb$^{-1}$. KLOE is now performing an
analysis using 2002 data ($\sim 240$ pb$^{-1}$)~\cite{sm}, for which a total systematic error 
(experimental and theoretical) $<1\%$ is expected. The acceptance cuts of this analysis
will be unchanged; the improvement will be due to
better and more stable running conditions as well as due to modification in the
online and offline environment, which will result in lower systematic errors associated
to the trigger and background-filter efficiencies. Concerning theory, a new version
of BABAYAGA is available~\cite{babayaga}, which allows to reduce the luminosity error by about a factor 2. The final
goal of the analysis is a measurement of R, which requires that 
$e^+e^-\to \mu^+\mu^-\gamma$ events are selected with equally high precision as 
$\pi^+\pi^-\gamma$ events .
\begin{figure}[t]
\epsfig{file=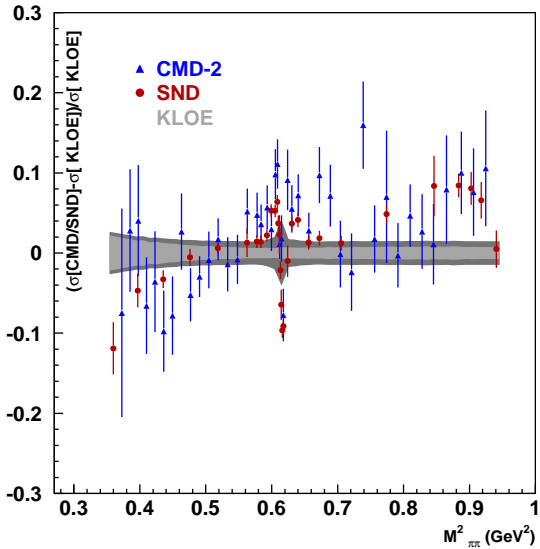, width=8cm}
\vspace{-0.75cm}
\caption{\small{Relative difference of the pion formfactor measurements from the experiments 
CMD-2 (triangles) and SND (circles), relative to the KLOE measurement. The KLOE data points have been 
interpolated 
and the statistical (light grey) and the systematic (dark grey) error bands are shown in the plot.}}
\label{diff}
\end{figure}
\subsection{Large angle analysis}
The analysis described above, in which the ISR-photon is emitted at small polar angles, does
not allow to cover the threshold region $M_{\pi\pi}^2 < 0.35 {\rm GeV}^2$, since in this kinematical
region the two pions are emitted essentially back-to-back to the ISR-photon and hence cannot be
detected simultaneously in the fiducial volume 
defined for the pion tracks $50^{\rm o}<\Theta_\pi<130^{\rm o}$~\cite{dl}. In order to
measure the pion formfactor at threshold, KLOE is now performing a complementary
analysis, in which the ISR-photon is tagged at large polar angles $50^{\rm o}<\Theta_\gamma<130^{\rm o}$.
Due to the $1/s^2$ dependence in the dispersion integral for $a_\mu^{\rm hadr}$,
the low mass region of the two-pion cross section is actually giving a $\sim 20\%$ contribution
to the total integral and hence an improved determination of the cross section at threshold is 
needed. 
\\
At large photon angles background from $\phi\to\pi^+\pi^-\pi^0$ is huge and dedicated selection 
cuts, like a cut on the angle between the missing momentum and the tagged photon direction,
as well as a kinematic fit in the background hypothesis with a cut on $\chi^2_{\pi\pi\pi}$
are needed to suppress this contribution. Moreover, irreducible background from events with 
the same $\pi^+\pi^-\gamma$ final state is not negligible anymore at large photon angles. 
This background category, which has to be subtracted relying on Monte-Carlo prediction, is
given by FSR-events and by the $\phi$ radiative decay into the scalar $f_0(980)$ with 
$f_0(980) \to \pi^+\pi^-$. A possible model dependence of FSR (the model of scalar QED is used
in PHOKHARA) and of the description of the scalar $f_0(980)\gamma$ amplitude, can be tested
by means of the forward-backward asymmetry.
In a recent KLOE publication~\cite{cesare} good agreement between data and simulation has been found
for the forward-backward asymmetry, setting upper limits for the systematic errors associated with 
these model uncertainties.
\\
\\
The main limitation for the measurement of the pion formfactor at threshold will arise from
the background channels $\phi\to\pi^+\pi^-\pi^0$ and $\phi\to f_0(980)\gamma \to \pi^+\pi^-\gamma$.
In order
to further reduce the systematic errors associated to these channels, 
the DA$\Phi$NE collider has taken data 
off-resonance at a center-of-mass energy of $\sqrt{s}=1.00$ GeV in its last KLOE run
(December 2005 to March 2006, $250 {\rm pb}^{-1}$ integrated luminosity). 
The off-peak analysis will allow a considerably improved determination of
the threshold region. Moreover, together with the data taken on-peak it will be possible to study 
the interference of the $f_0(980)$ amplitude with FSR.

\section{THE RADIATIVE RETURN AT PEP-II WITH THE BABAR DETECTOR}
The BaBar experiment has previously published results of 
the exclusive $\pi^+\pi^-\pi^0$, $\pi^+\pi^-\pi^+\pi^-$, $K^+K^-\pi^+\pi^-$ and
$K^+K^-K^+K^-$ final states with better coverage than all previous
experiments and comparable or better precision, using 89 fb$^{-1}$ of data.
New results are presented here for final states with 
$6$ hadrons and for the timelike proton-antiproton formfactor.
The new results contain a larger data set of $\sim 240 {\rm pb}^{-1}$. 
The mass range of interest in all these measurements 
$M_{\rm hadr}<4.5$ GeV is rather
distant from the center-of-mass energy of PEP-II ($\sqrt{s}=m_{\Upsilon(4S)}=10.58$ GeV),
requiring hence a high-energetic ISR-photon with an energy in the
center-of-mass system of $E_\gamma^*>3$ GeV. 
Such an event signature separates ISR-events from background from the
$\Upsilon(4s)$-resonance (B-decays and their consecutive decay products), 
since photons originating from these events are in general much less-energetic. 
The event signature at the B-factory also implies the necessity for 
photon tagging in order to have high acceptance for the hadronic system.
The mass resolution is improved by performing a kinematic fit in the signal hypothesis, 
requiring constraints on four-momentum and (in case of neutral pions) of the $\pi^0$-mass.
A cut on the $\chi^2$ of the kinematic fit is the main tool for background subtraction and
the shape of the $\chi^2$-distribution is used with a sideband-method to obtain the 
residual background after all cuts.\\
\begin{figure}
\begin{center}
\epsfig{file=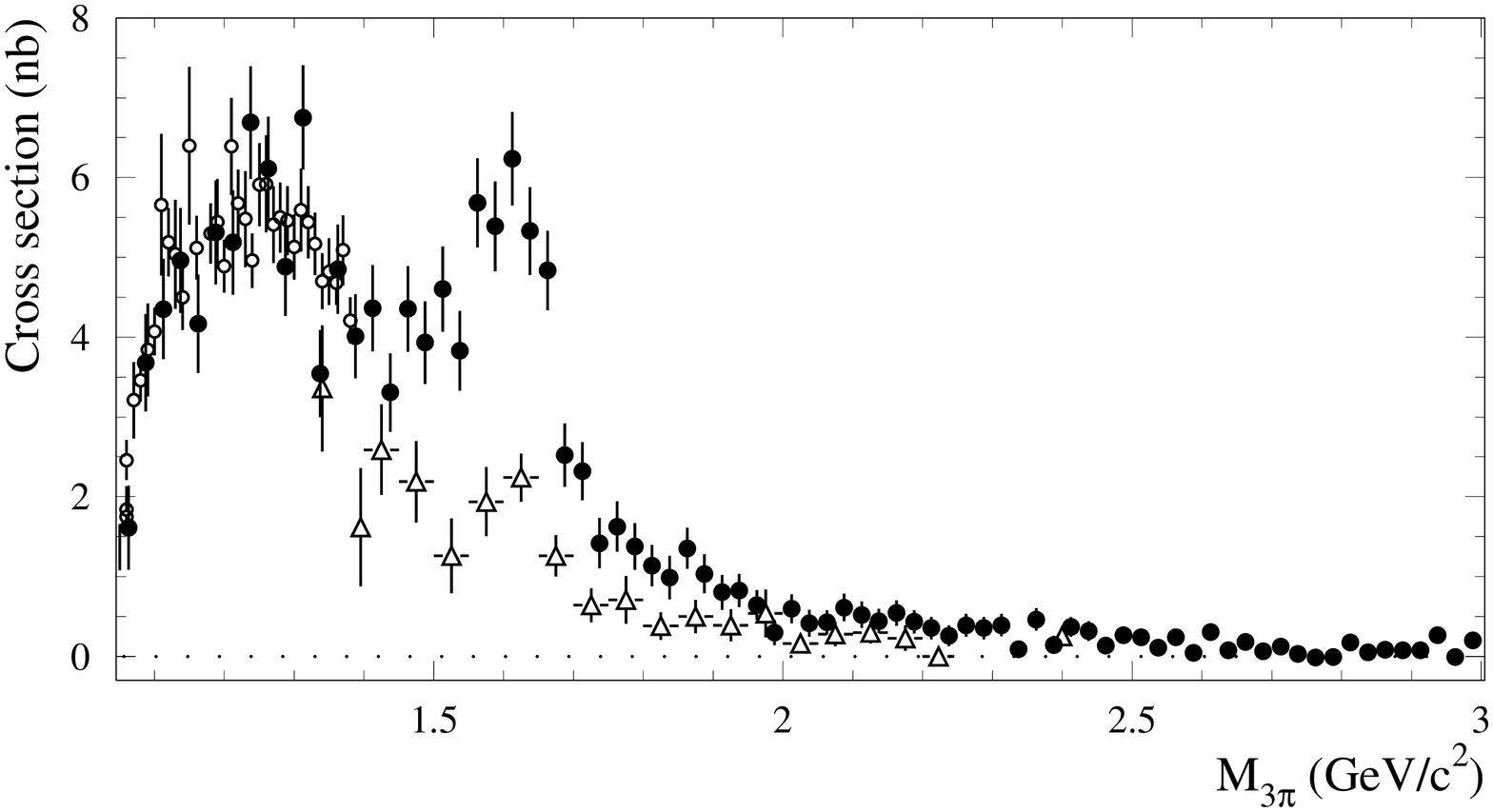, width=7.5cm, height=5cm}
\end{center}
\vspace{-1.0cm}
\caption{\small{BaBar measurement (filled circles) of the cross section
$e^+e^-\to \pi^+\pi^-\pi^0$, compared to previous measurements from SND (open circles)
and DM2 (open triangles).}}
\label{fig.3pi}
\end{figure}
In this paper we report not only on the cross section measurements, but we also
present studies of the internal structures. For all channels discussed here the
$J/\psi$ signals were used to extract the associated branching fractions and in
some cases this has been done also for the $\psi(2S)$ signals.
BaBar has published $10$ $J/\psi$ and $3$ $\psi(2S)$ branching ratios; in $4$ cases there
has been no previous measurements, in $5$ cases the measurements are better
than the current world average. 
In the following we give a brief overview of the individual channels 
published so far by BaBar with special emphasis on the more recent analyses of
$6$ Hadrons and $e^+e^-\to p\bar{p}$.

\subsection{$e^+e^-\to 3$ Pions} 
The $\pi^+\pi^-\pi^0$ mass spectrum has been measured from $1.05$ GeV up to the $J/\psi$ mass region 
with a systematic error of $\sim 5\%$ below $2.5$ GeV and up to $\sim 20\%$ at higher masses~\cite{3piref}. 
The spectrum is dominated by the $\omega$, $\phi$ and $J/\psi$ resonances. The
BaBar measurement could improve significantly on the world's knowledge of the 
excited $\omega$ states. Fig.~\ref{fig.3pi} shows the BaBar data points 
together with previous measurements in the energy range relevant 
for the exited states $\omega^\prime$ and $\omega^{\prime\prime}$.
A clear disagreement with DM2~\cite{dm23pi} is seen above $1.4$ GeV. The BaBar spectrum has been fitted 
up to $1.8$ GeV and the following results for the masses and widthes of the
$\omega^\prime$ and $\omega^{\prime\prime}$ have been found\footnote{in this fit 
$M_{\omega,\phi}$ and the phases between $\omega$, $\phi$ with respect to
$\omega^\prime$, $\omega^{\prime\prime}$ are fixed.}:
$M(\omega^\prime)=(1350\pm20\pm20)$ MeV, $\Gamma(\omega^\prime)=(450\pm70\pm70)$ MeV,
$M(\omega^{\prime\prime})=(1660\pm10\pm2)$ MeV, 
$\Gamma(\omega^{\prime\prime})=(230\pm30\pm20)$ MeV.
\begin{figure}[t]
\epsfig{file=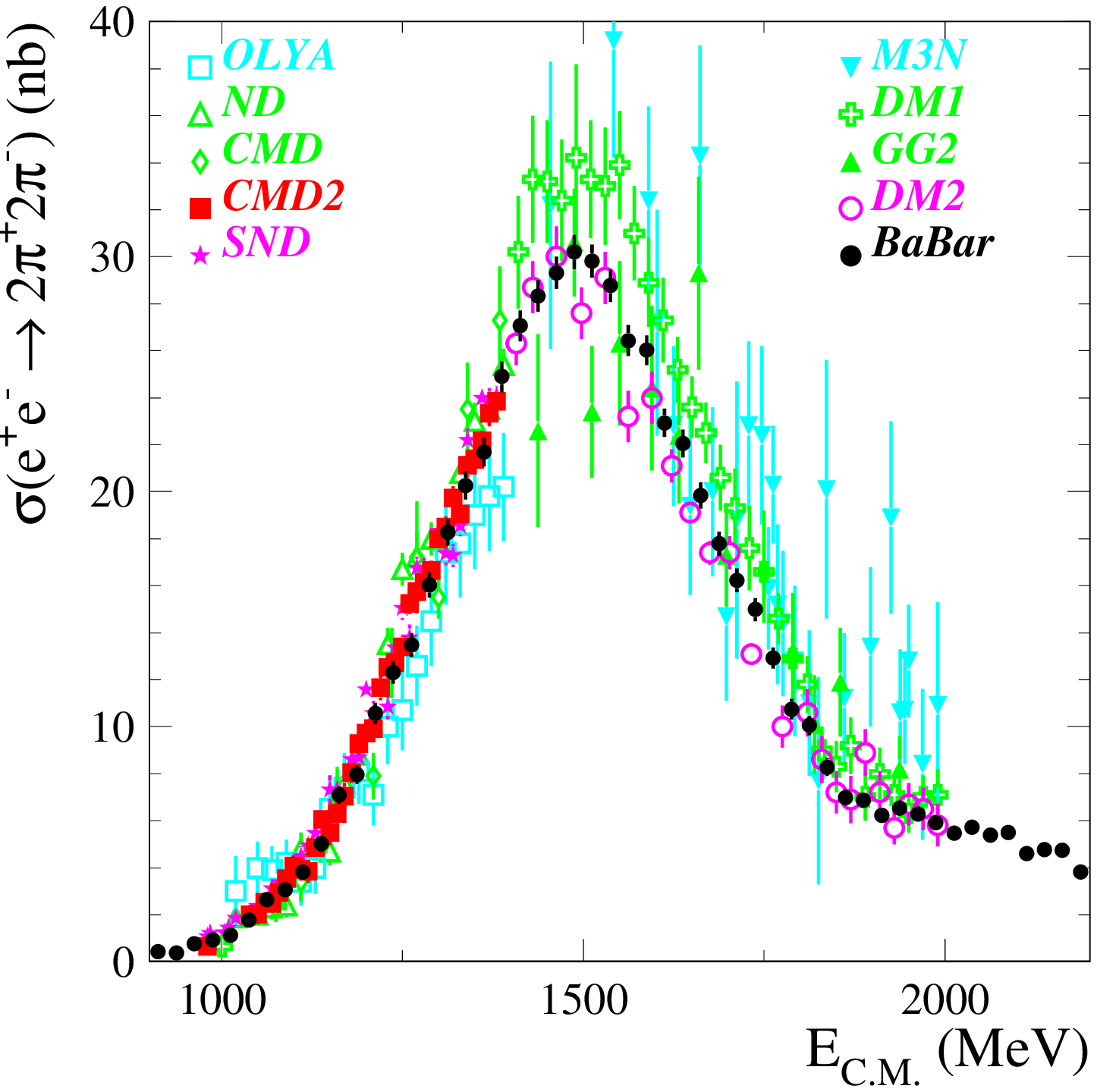, width=6.cm}
\epsfig{file=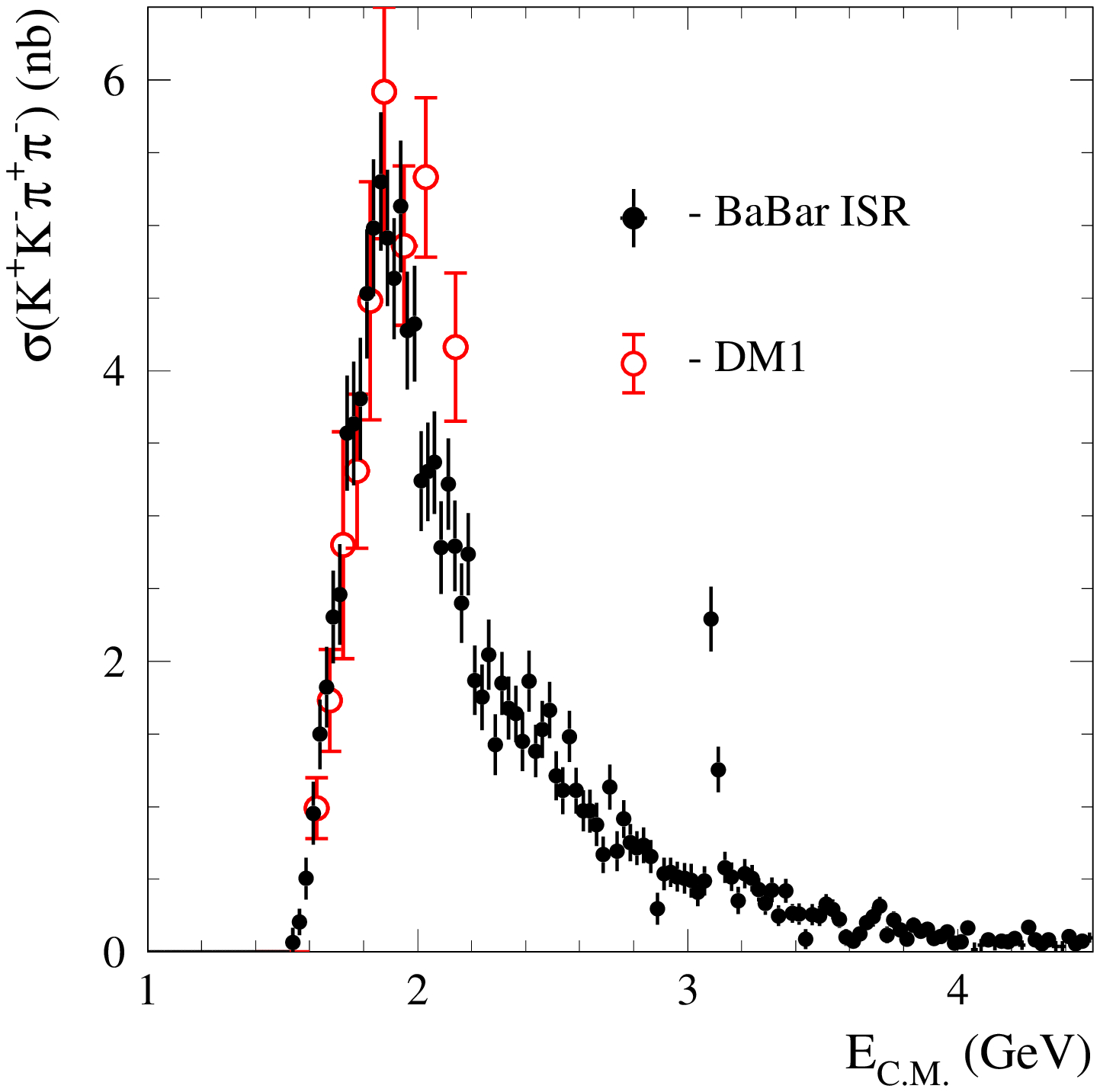, width=6.cm}
\vspace{-1.0cm}
\caption{\small{The energy dependence of the $e^+e^-\to \pi^+\pi^-\pi^+\pi^-$ (upper plot)
and the $e^+e^-\to K^+K^-\pi^+\pi^-$ (lower plot) cross section obtained by BaBar (filled circles) 
by radiative return in comparison with previous data.}}
\label{fig.4pi}
\end{figure}

\subsection{$e^+e^-\to 4$ Hadrons}
The $\pi^+\pi^-\pi^+\pi^-$, $K^+K^-\pi^+\pi^-$ and $K^+K^-K^+K^-$ exclusive final states 
have been measured from theshold up to $4.5$ GeV with systematic errors of $5\%$, $15\%$ and
$25\%$, respectively~\cite{4piref}. Fig.~\ref{fig.4pi} shows the mass distribution of the events satisfying the
$\pi^+\pi^-\pi^+\pi^-$ and $K^+K^-\pi^+\pi^-$ kinematics as well as $K/\pi$ identification 
constraints. We identify that the radiative return technique at BaBar allows to cover a 
wide energy range in one single experiment. The $K^+K^-K^+K^-$ measurement is the first measurement ever.
Background is relatively low for all channels under study 
(e.g. few percent at $1.5$ GeV for $\pi^+\pi^-\pi^+\pi^-$) 
and is dominated by ISR-events of higher multiplicities and of continuum non-ISR 
events at higher masses. 
\\
The $\pi^+\pi^-\pi^+\pi^-$ final state is dominated by the two-body $a_1(1260)\pi$ intermediate state;
the $K^+K^-\pi^+\pi^-$ final state shows no significant two-body states, but rich three-body
structure, including $K^*(890)K\pi$, $\phi\pi\pi$, $\rho K K$ and $K_2^*(1430)K\pi$. 
\begin{figure}[t]
\epsfig{file=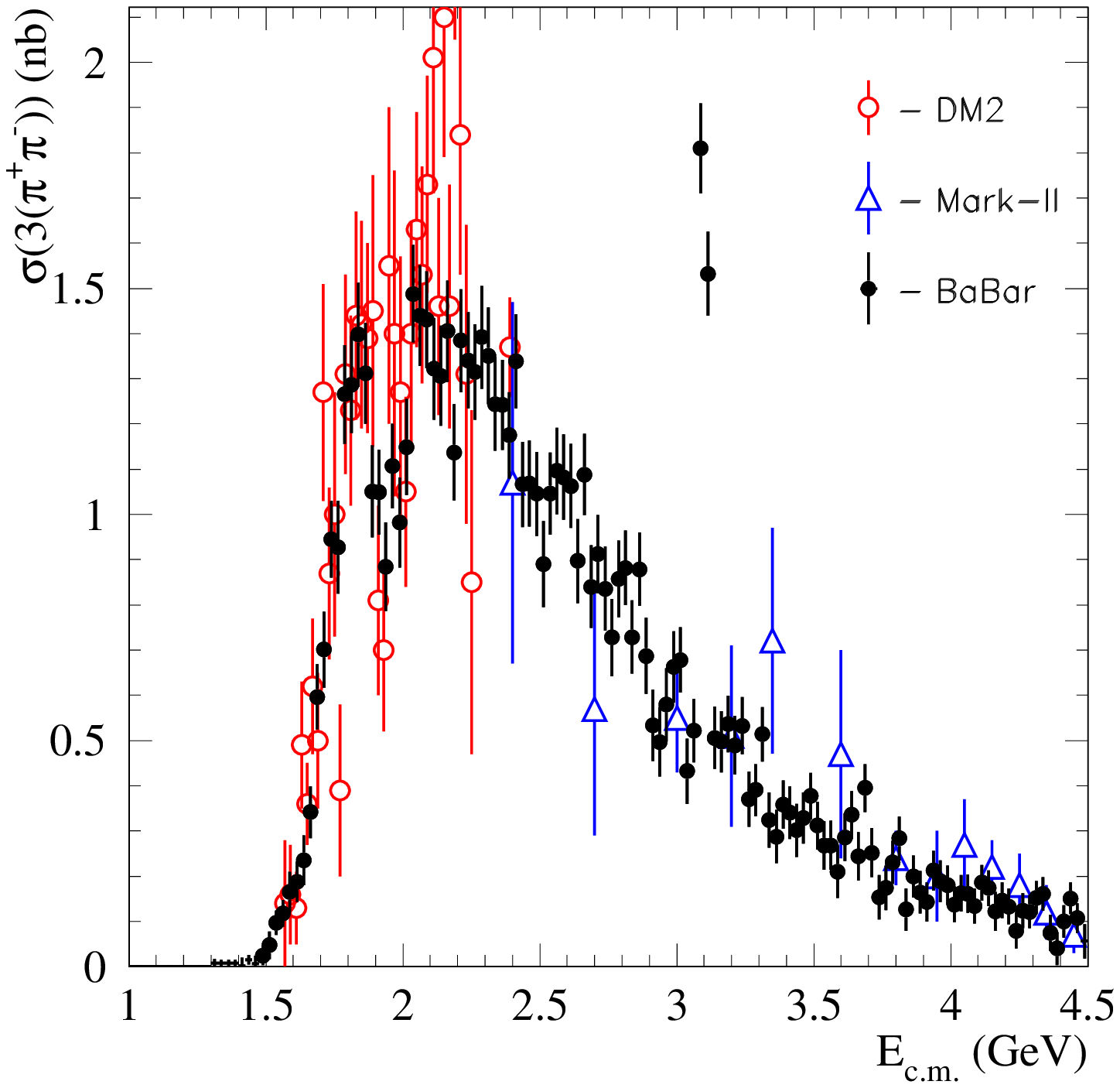, width=6.cm}
\epsfig{file=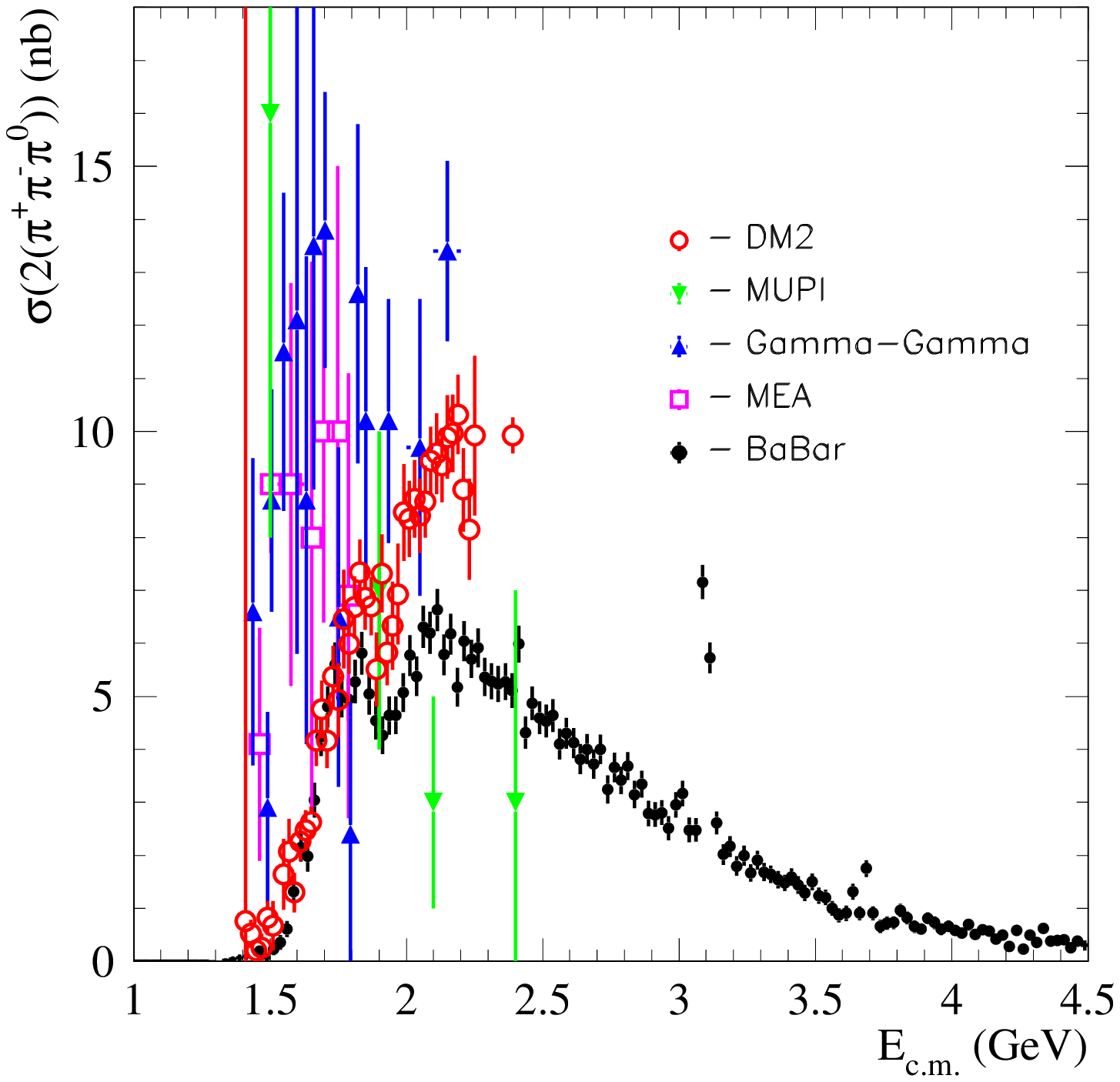, width=6.cm}
\vspace{-1.0cm}
\caption{\small{The energy dependence of the $e^+e^-\to 3(\pi^+\pi^-)$ (upper plot)
and the $e^+e^-\to 2(\pi^+\pi^-)2\pi^0$ (lower plot) cross section obtained by BaBar (filled circles) 
by radiative return in comparison with previous data.}}
\label{fig.6pi}
\end{figure}  
\begin{figure}[t]
\epsfig{file=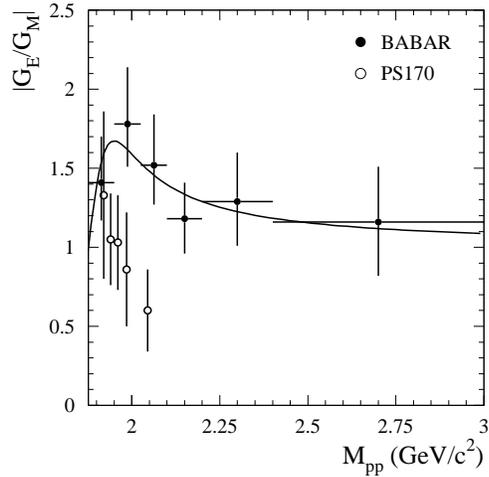, width=6.8cm}
\vspace{-1.0cm}
\caption{\small{BaBar measurement of the ratio of the electric and magnetic formfactor 
describing the cross section $e^+e^-\to p\bar{p}$. Filled circles depict BaBar
data, the curve is a fit result. Open circles show data from the experiment PS170.}}
\label{fig.gegm}
\end{figure} 
\subsection{$e^+e^-\to 6$ Hadrons} 
The $6$-Hadrons final state has been measured in the exclusive channels 
$3(\pi^+\pi^-)$, $2(\pi^+\pi^-)2\pi^0$ and $K^+K^-2(\pi^+\pi^-)$~\cite{6piref}. 
The cross section in the last case has never been measured before; the precision
in the first two cases is $\sim 20\%$, which is a large improvement with respect to existing data.  
Again, the entire energy range from threshold up to $4.5$ GeV is measured
in one single experiment.
\\
The distributions for the final states $3(\pi^+\pi^-)$ and $2(\pi^+\pi^-)2\pi^0$
are shown in fig.~\ref{fig.6pi}. A clear dip is visible at about $1.9$ GeV in both 
modes. A similar feature was already seen 
by FOCUS~\cite{focus}  in the diffractive photoproduction of six charged pions.
The spectra are fitted using the sum of a Breit-Wigner resonance function 
and a Jacob-Slansky continuum shape. For the
$3(\pi^+\pi^-)$ ($2(\pi^+\pi^-)2\pi^0$) mode, we obtain values of $1880 \pm 30$ MeV
($1860 \pm 20$ MeV) for the resonance peak, $130 \pm 30$ MeV
($160 \pm 20$ MeV) for the resonance width  and $21^{\rm o} \pm 14^{\rm o}$
($-3^{\rm o} \pm 15^{\rm o}$) for the phase shift between the resonance and continuum.
The width values differ significantly from the FOCUS result of $29 \pm 14$ MeV.
\\
Also the substructures of the individual modes have been investigated.
There is surprisinlgly little substructure in the $3(\pi^+\pi^-)$ channel. The
spectrum can be described by a simulation with one $\rho^0$ and four pions 
distributed according to phase space. The $2(\pi^+\pi^-)2\pi^0$ final state
shows a similar cross section but a much more complex internal structure.
BaBar observes signals for $\rho^0$, $\rho^\pm$, $\omega$ and $\eta$, and
a substantial contribution from the two-body $\omega\eta$
intermediate state, which appears to be resonant. Additional modes 
containing more $\pi^0$ are under study. Also the $K^+K^-2(\pi^+\pi^-)$
channel shows an interesting substructure with a rather weak $\phi$ 
contribution, but a strong $K^*(890)$.

\subsection{$e^+e^-\to p \bar{p}$} 
BaBar has also performed a measurement of the $e^+e^-\to p\bar{p}$ cross section~\cite{ppbarref}.
The experimental challenge in this case is to fight the $KK$, $\pi\pi$ and $\mu\mu$
two-body background, which has a cross section approximately a factor $10-100$ higher.
Apart from the good momentum resolution of the detector, BaBar has excellent particle identifacation
capabilities, which allow a clean selection of protons; $\sim 4000$ $p\bar{p}$ 
events are selected from a data sample of $\sim 240 {\rm pb}^{-1}$.
The timelike proton-antiproton form factor is parametrized by an electric 
and a magnetic formfactor $G_E$, $G_M$:
\begin{eqnarray*}
\sigma_{e^+e^- \to p \bar{p}}(s) & = & \frac{4\pi\alpha^2 C}{3s} \sqrt{1 - \frac{2m_p^2}{s}} \\
                                 & {\rm x} & (|G_M(s)|^2 + \frac{2m_p^2}{s} |G_E(s)|^2),
\end{eqnarray*}
\begin{figure}[t]
\epsfig{file=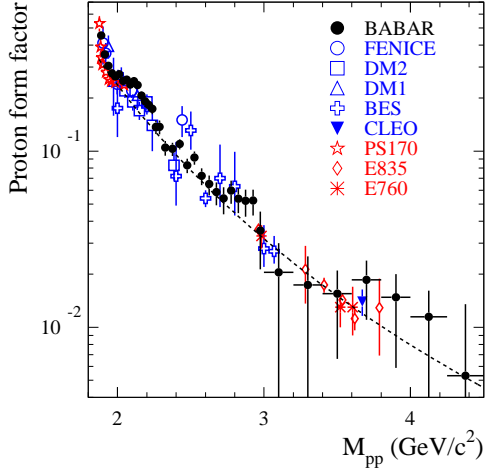, width=6.8cm}
\vspace{-1.0cm}
\caption{\small{The $e^+e^-\to p\bar{p}$ cross section measured by BaBar (filled circles) 
in comparison with data from other $e^+e^-$ colliders.}}
\label{fig.ppbar}
\end{figure} 
where the factor $C$ accounts for the Coulomb interaction of the final state particles.
The proton helicity angle $\theta_p$ in the $p\bar{p}$ rest frame can be used to
separate the $|G_E|^2$ and $|G_M|^2$ terms. Their respective variations are approximately 
$\sim \sin^2\theta_p$ and $\sim (1+\cos^2\theta_p)$.
By fitting the $\cos\theta_p$ distribution to a sum of the two terms,
the ratio $|G_E/G_M|$ can be extracted. This is done separately in six bins 
of $M_{p\bar{p}}$. The results are shown in fig.~\ref{fig.gegm}, and disagree significantly
with previous measurements from LEAR~\cite{lear} close to threshold. At larger values of
$M_{p\bar{p}}$ the BaBar measurement finds $|G_E/G_M| \approx 1$.
\\
In order to compare the cross section measurement with previous data ($e^+e^-$ and $p\bar{p}$ 
experiments), the {\it effective} form factor is introduced: 
$G = \sqrt{|G_E|^2 + 2m_p^2/s |G_M|^2}$. The BaBar measurement of 
$G$ is in good agreement with existing results, as can be seen in fig.~\ref{fig.ppbar}. The structure of
the formfactor is rather complicated; we make the following observations:
(i) BaBar confirms an increase of $G$ towards threshold as seen before by other experiments; 
(ii) two sharp drops of the spectrum at $M_{p\bar{p}}=2.25$ and $3.0$ GeV are observed; 
(iii) data at large values
$M_{p\bar{p}}>3$ GeV is in good agreement with the prediction from perturbative QCD. 

\section{CONCLUSIONS}
At the particle factories DA$\Phi$NE and PEP-II the radiative return method allows precision
measurements of the hadronic cross section, which are  
of utmost importance for an improved determination
of the hadronic contribution of the muon anomaly $a_\mu^{\rm hadr}$ 
and of the running fine structure constant $\alpha(M_Z^2)$.
At DA$\Phi$NE the KLOE experiment has measured
the pion formfactor with a precision of $1.3\%$. An update of the analysis will lead to a
further reduction of the systematic error and a normalization to radiative muon pairs
is foreseeen. Moreover, KLOE has collected
data with the collider running off-resonance, which will allow an improved determination
of the threshold mass region. \\
At the B-factory PEP-II a program to measure all exclusive hadronic channels from theshold
up to $4.5$ GeV is underway. In addition to the published results presented in this paper,
there are ongoing analyses for further final states, including the important two-pion-
channel. 

\end{document}